# Epitaxial Growth of Ultra-smooth δ-NbN Thin Films on TiN-Buffered Sapphire by Room-Temperature Sputtering

Swagata Bhunia[1,2,*], Aakash Shandilya[2], Sounak Samanta[2], Bikash C Barik[2], Soumyadip Chatterjee[1], Parushottam Majhi[2], Siddarth Rastogi[2], Kantimay Das Gupta[2], Suddhasatta Mahapatra[2,*] and Apurba Laha[1,*]

1. Department of Electrical Engineering, Indian Institute of Technology, Bombay, Powai, Mumbai-400076.
2. Department of Physics, Indian Institute of Technology, Bombay, Powai, Mumbai-400076.

- Corresponding authors email: suddho@phy.iitb.ac.in, laha@ee.iitb.ac.in, swagatabhunia@gmail.com

Abstract: The δ phase of Niobium Nitride (NbN) is a promising superconducting material, which is chemically stable and shares lattice compatibility with conventional III-Nitride semiconductors. Due to a high critical temperature ($T_c$) and a high critical (magnetic) field (Hc), NbN is much-coveted for a diverse set of applications spanning from single photon detectors, and hot-electron bolometers to quantum computing architectures using superconducting circuits. However, synthesizing high-quality epitaxial films of phase pure and stoichiometric δ-NbN in a cost-effective manner, is challenging. In this study, we investigate the epitaxial growth of single crystalline δ-NbN on TiN-buffered c-sapphire ($Al_2O_3$) substrates by sputtering at room temperature. For these films, we demonstrate a surface-roughness in picometer-scale, the lowest reported till date. The critical temperature ($T_c$) of the epitaxial δ-NbN films was observed to decrease with the insertion of the TiN buffer layer, tentatively attributable to the leakage of Cooper pairs, due to the proximity effect. TiN and NbN layer behave as a bilayer system, wherein Cooper-pair leakage is facilitated by the absence of any oxide interlayer. Consequently, $T_c$ reduces with increasing thickness of the TiN layer.

In recent years, the superconducting transition-metal nitride, niobium nitride (NbN), has been extensively studied, owing to its potential use in the development of a wide variety of (opto) electronic devices, such as Josephson junctions,[1-4] single-photon detectors,[5] hot-electron bolometers,[6] and kinetic inductance detectors.[7] NbN is chemically stable and exhibits a high critical temperature ($T_c$),[8, 9] a large critical magnetic field,[10] and a short coherence length.[11] Moreover, due to its compatibility with the conventional III-nitride semiconductors, NbN can be monolithically integrated to next-generation nitride-based semiconductor-superconductor devices. Examples include, the AlGaN/GaN high-electron-mobility transistor (HEMT) grown on NbN thin films for advanced microwave amplifiers,[12] Josephson junctions based on NbN/AlN/NbN heterostructures for superconducting qubits,[4] and superconducting quantum interference devices (SQUIDs),[13, 14] and NbN superconducting nanowires integrated with InGaN QDs for single photon detection. Such devices have crucial applications in the fields of magnetic sensing, quantum computing and quantum secure communications. However, their performance heavily relies on the phase, crystallinity of NbN, and the quality of NbN/AlN interface. It is observed that among all the known phases (β, γ, ε, δ),[15-18] the cubic phase (δ) of NbN is the preferable one, since it shows the highest $T_c$ (17 K),[18, 19] and the minimum lattice mismatch with AlN (0.2% and GaN 2.7%).[20] However, focused research efforts are required to establish economical and commercially viable methods of depositing phase-pure epitaxial films of NbN, on low cost commensurate substrates or suitable buffer layer grown on them.

Due to very small lattice mismatch with NbN, (001) oriented AlN buffer layers grown by using metal-organic chemical vapor deposition (MOCVD) or molecular beam epitaxy (MBE) on c-plane sapphire substrates, is the ideal candidate for the growth of superconducting nitride. However, the processing cost in this case is very high.[2, 21] Moreover, the high melting point of Nb (~ 2600 ˚C) makes MBE a rather challenging option.[22] In contrast, sputtering of a Nb target is straightforward and cost-effective. Therefore, it is desirable to select a material for the buffer layer that not only has a low lattice mismatch with δ-NbN but can also be deposited using sputtering. Titanium nitride (TiN) is one such candidate, which can be sputter deposited to obtain highly crystalline thin films on c-sapphire and has a relatively small lattice mismatch (3.5%) with NbN.[23]

In this study, we demonstrate epitaxial growth of δ-NbN thin films with sub-nanometer-scale surface roughness by sputter deposition at room-temperature, on TiN (001)/c-sapphire virtual substrates. Our results indicate that a low thickness of the TiN layer reduces the surface roughness of the NbN layer, tentatively due to the suppression of strain relaxation, and the concomitant absence of resulting threading dislocations. Furthermore, the temperature-dependent four probe measurements reveal a lowering of $T_c$, as the thickness of the TiN layer increases. A detailed analysis indicates that NbN and TiN act as a bilayer system that leads to the leakage of Cooper-pairs into the TiN layer.

## Results and Discussion:

Figure 1(a) shows the XRD wide-angle ω-2θ diffractograms recorded from samples A, B, C, D, and E. For all samples, the diffraction peaks observed at 2θ = 20.49˚, 41.67˚, and 64.49˚, correspond to the (003), (006) and (009) planes of sapphire, respectively. For sample A, the additional peak observed at 2θ = 35.60˚ may be attributed to the (111) planes of δ-NbN. The absence of diffraction peaks corresponding to other phases of NbN, indicates that the room-temperature-deposited NbN thin-film is phase-pure δ-NbN, with a cubic crystal structure. Although it is single crystalline, the broad and very low intensity of the (111) reflection suggests a relatively poor crystalline quality. For sample B on the other hand, together with the δ-NbN an additional peak is detected at 2θ = 36.71˚. This corresponds to the (111) planes of TiN. Since peaks from other TiN phases are not observed in the diffractogram, we infer that the deposited TiN is also cubic. The XRR analysis reveals that the approximate thicknesses of the TiN and NbN layers are 72 nm, and 30 nm, respectively, in sample B. Interestingly, the insertion of the TiN buffer layer has led to a significant increase in the NbN (111) peak intensity, clearly indicating an improvement of the crystal quality of the latter. We attribute this improvement to the lower lattice mismatch between NbN and TiN compared to that between NbN and sapphire, which therefore yields a lower density of misfit dislocations.

**Table 1.** Details of the samples grown on sapphire (001).

| Sample ID | Deposition time of TiN | Deposition time of NbN | Thickness of NbN | Thickness of TiN |
|---|---|---|---|---|
| sample A | - | | | - |
| sample B | 30 min | | | ~ 72 nm |
| sample C | 15 min | 25 min | ~ 30 nm | ~ 36 nm |
| sample D | 4 min 10 s | | | ~ 10 nm |
| sample E | 2 min 5 s | | | ~ 5 nm |

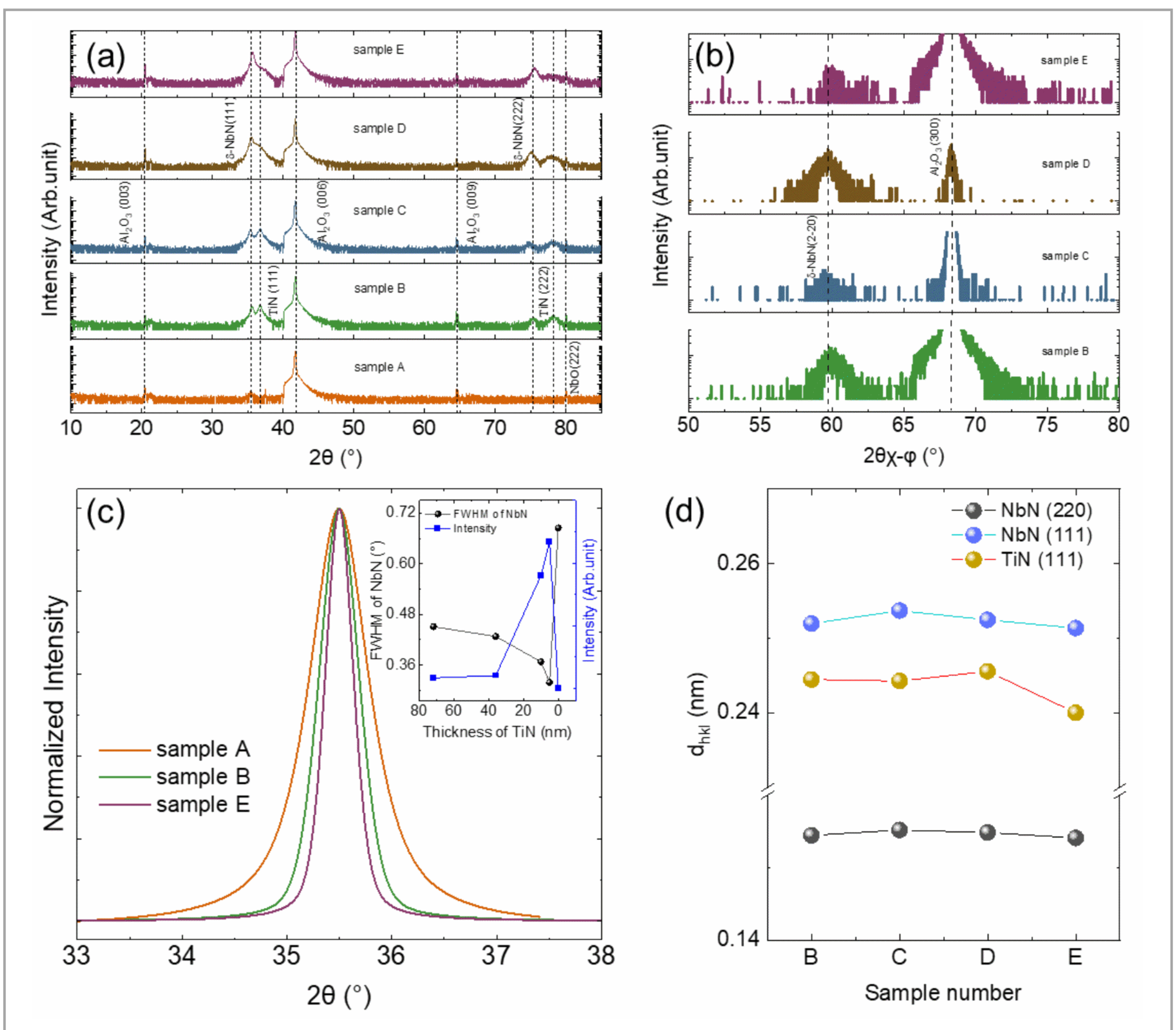

**Figure 1.** (a) Out-of-plane XRD and (b) in-plane XRD of the sample without a TiN buffer (sample A) and with a TiN buffer (sample B to sample E). (c) NbN (111) peaks, extracted from fitting the out of plane diffractogram of the respective samples; the inset shows the FWHM and NbN intensity for all samples. (d) Interplanar spacing calculated from the out of plane and in plane XRD diffractogram of NbN and out of plane XRD diffractogram of TiN.

Thickness of TiN layer decreases progressively from sample B to sample E. Concomitantly, the normalized-peak-intensity (normalized to substrate) of the NbN (111) reflection is observed to increase continuously (inset of Fig. 1(c)). Furthermore, the higher order reflections from the {111} family of planes of both NbN and TiN planes are also observed at $2\theta = 75.28^\circ$ and $78.25^\circ$, respectively for all of these samples. This further supports the inference of improved crystallinity of the NbN film, with the inclusion of the TiN buffer layer. The peak at 2θ = 80.01˚, seen in all samples, corresponds to the (222) planes of NbO, indicating that a few monolayers at the top of the NbN deposit are oxidized. While a (111)-oriented growth of NbN is established by the out of plane XRD, in-plane $2\theta_{\chi}$-Φ diffractograms were recorded for all samples, to investigate the epitaxiality of the films (Fig. 1(b)). The peaks observed at 59.63˚, and 68.23˚ correspond to the (2-20) planes of δ-NbN and (300) planes of sapphire, respectively. This indicates that δ-NbN grown on TiN/sapphire is epitaxial in nature, despite being deposited at room temperature. The epitaxial relationship of δ-NbN with sapphire can be expressed as NbN [1-10] ‖ $Al_2O_3$ [100] and NbN [111] ‖ $Al_2O_3$[001]. To further evaluate the epitaxial quality, the NbN (111) peak of the ω-2θ diffractograms were fitted with the pseudo-Voigt function and the fitted peaks are shown in Fig. 1(c), for a few samples. The inset of Fig. 1(c) shows the corresponding FWHM values and the normalized-peak intensity. A comparison of the values for samples A and E reveals a two-fold reduction in FWHM with the inclusion of the ~ 5 nm-thick-TiN buffer layer. However, as the TiN layer thickness increases, the FWHM increases further. The NbN peak intensity follows the opposite trend as that of the FWHM, as expected. We corroborate this observation to the strain state of the TiN buffer layer. While the thinner TiN layer of sample E is pseudomorphically strained, that of sample B is tentatively partially relaxed. This is expected to introduce threading dislocations in the buffer layer, which percolate into the NbN layer on top. To understand the strain-relaxation behavior further, the inter-planar spacing of the (111) and the (220) planes of NbN, and the (111) planes TiN were plotted, as shown in Fig. 1(d). The TiN (111) inter-planar spacing slightly increases with decreasing TiN thickness for samples B to D, but decreases again for sample E. The spacings of the NbN (111) planes show a similar behavior; however, no variation is observed for the spacings of NbN (220) planes. This implies that both TiN and NbN layers are strained in samples B to D. However, for sample E, the TiN (111) planes appear to be relaxed as indicated by the reduction in the (111) inter-planar spacing. This may be due to an error in determining the exact Bragg angles of TiN (111) planes, as the diffraction peak is already broadened.

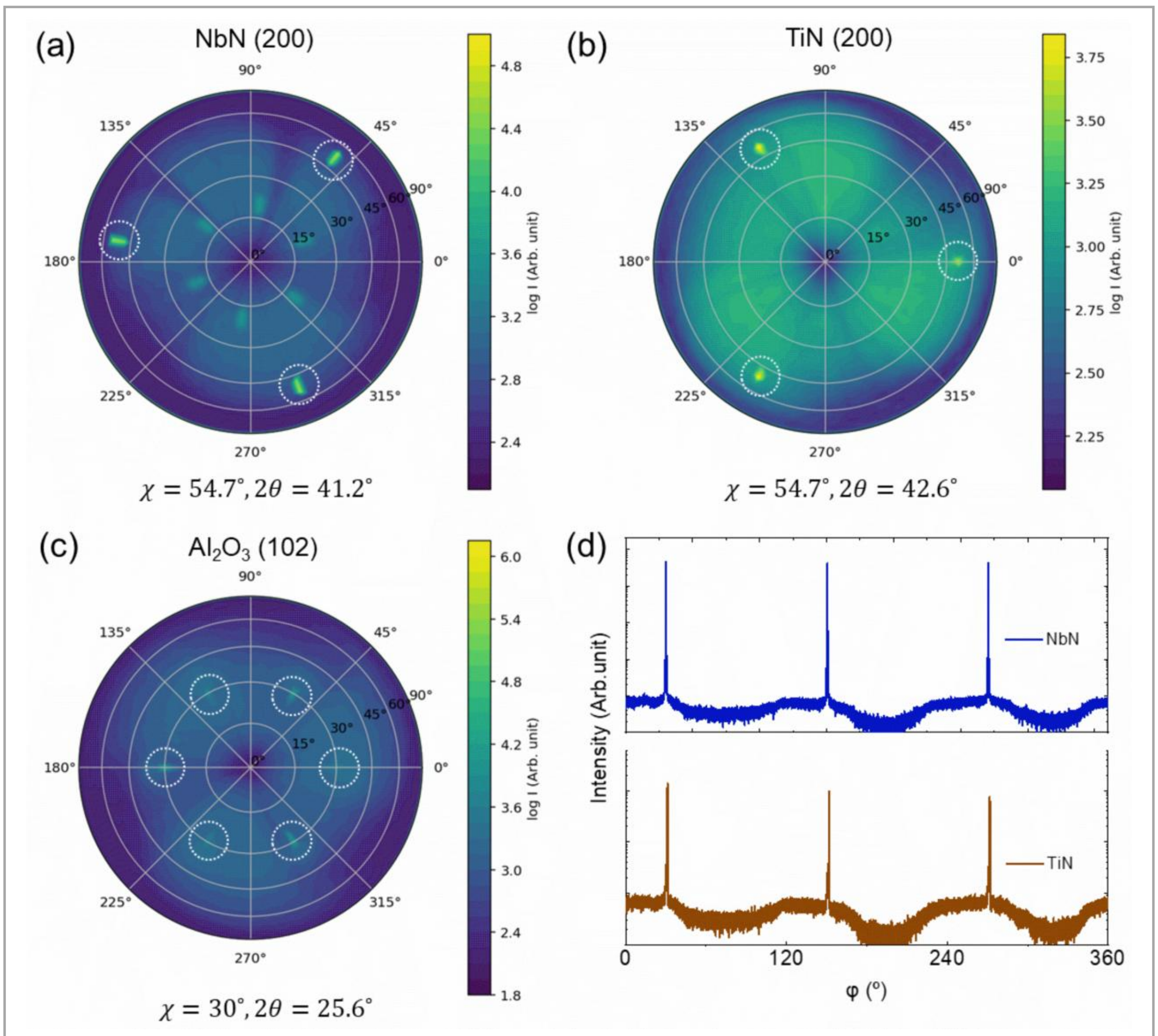

**Figure 2.** (a-c) Pole-figures of the NbN (200), TiN (200) and $Al_2O_3$ (102) planes, respectively. The poles in the figures are shown by white dotted lines. (d) $\phi$-scan of the (200) planes of NbN and TiN, showing three-fold symmetry with no rotational mis-orientation.

X-ray pole figure recorded from sample B, for the Bragg angle corresponding to the NbN (200) reflections ($2\theta = 41.2^\circ$) is shown in Fig. 2(a). The figure shows three clear diffraction spots at $\chi = 54.7^\circ$ (marked by the white dotted circles), corresponding to the (200) planes of NbN. This reflects the threefold symmetry of the {200} planes. In the same figure six low-intensity spots are also observed at $\chi = 23^\circ$, which stems from the (113) planes of sapphire. As the measurement was performed at $2\theta = 41.2^\circ$, which is very close to the Bragg angle of the sapphire (113) planes ($2\theta = 41.3^{\circ}$), the additional six poles may be attributed to the {113} reflections of sapphire. The presence of NbN (200) planes only at $\chi = 54.7^{\circ}$ indicates that the NbN domains are highly oriented, typical for epitaxial layer. The XRD pole figures corresponding to the Bragg angles of the TiN (200) and sapphire (104) planes ($2\theta = 42.6^{\circ}$ and $25.6^{\circ}$, respectively) are shown in Figs. 2(b, c). In Fig. 2(b), three poles are observed at $\chi = 54.7^{\circ}$ reflects the threefold symmetry of TiN {200} planes, thus confirming its cubic phase. On the other hand, Fig. 2(c) shows six poles (shown by dotted circles) at $\chi = 30^{\circ}$ due to sixfold symmetry of the (104) planes of sapphire. To further investigate whether there is any rotational misalignment between NbN and TiN crystal planes, $\phi$-scans of the {200} planes were performed, as shown in Fig. 2(d). Three sharp peaks are observed for both NbN and TiN, with no relative shift in $\phi$, indicating exact alignment between crystal planes of the nitrides.

The arrangement of the NbN unit cells, when grown directly on sapphire and on a TiN buffer layer, were simulated using VESTA software. The simulated unit cell of hexagonal sapphire, and cubic NbN and TiN crystals, are shown in Figs. 3(a)-(c), while their superpositions are presented in Fig. 3(d)-(e). Since the (111) planes of NbN and TiN are grown on the (001) plane of sapphire, the top-views of these planes are shown in Fig 3(b), Fig. 3(c) and Fig. 3(a), respectively. In the same figures, the out of plane and in-plane lattice constant of NbN, TiN and sapphire are also mentioned. When NbN is directly grown on sapphire, a large mismatch in the in-plane lattice constant is observed from the top-view image (Fig. 3(d)). However, when it is grown on TiN, the mismatch is significantly reduced, as seen in Fig. 3(e).

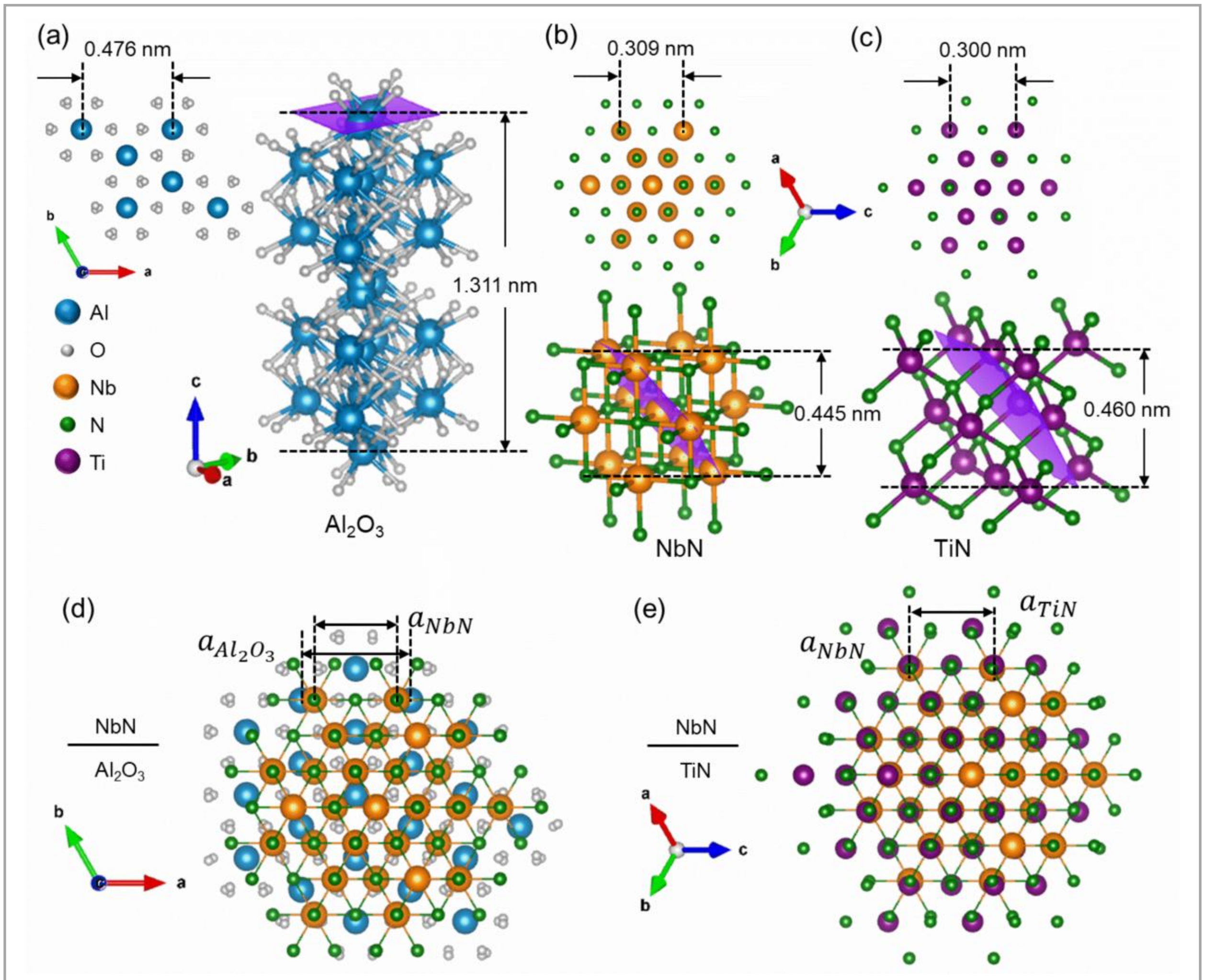

**Figure 3.** (a-c) Crystal structures of hexagonal $Al_2O_3$, cubic NbN and TiN, together with top view representations of the (001) plane of $Al_2O_3$ and the (111) planes of NbN and TiN. (d, e) top view images of NbN on $Al_2O_3$ and NbN on TiN, showing in-plane lattice mismatch of NbN and the respective underlying layer.

A smooth surface is desirable for better performance of device structure fabricated with superconducting films, as increased surface roughness can reduce the effective superconducting coherence length. To investigate the surface morphology of the films, AFM images were recorded on $10 \times 10\ \mu m^2$ area on each sample (Fig. 4). While the root-mean-square (RMS) roughness of sample A is 2.86 nm (Fig. 4(a)), due to the introduction of the TiN buffer layer in sample B, the RMS roughness is reduced to 0.344 nm. This 8-fold reduction in the surface roughness is remarkable, suggesting controlled strain relaxation, without island formation, tentatively due to the reduced lattice mismatch. This also reflects that the growth of the NbN films were not kinematically limited, despite sputter deposition at room-temperature.

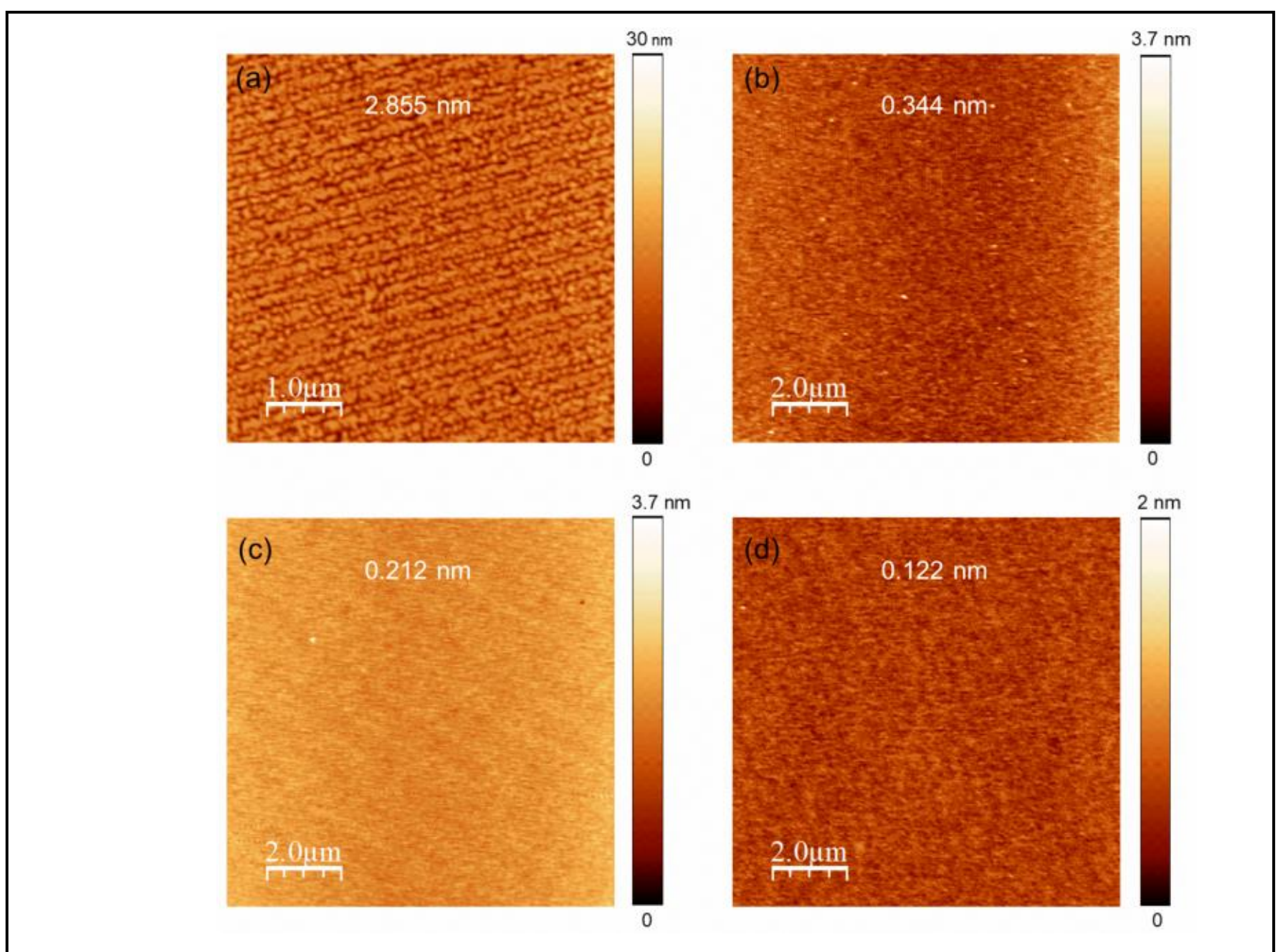

**Figure 4.** (a-d) AFM images of sample A, sample B, sample D, and sample E, respectively. (a) AFM scan over a $5 \times 5\ \mu m^2$ area while (b-d) show scans over a $10 \times 10\ \mu m^2$ area.

**Table 2.** Reported low surface roughness values of NbN films.

| NbN Thickness | Deposition Tool | Growth Temperature | Underlying Layer | Surface roughness |
|---|---|---|---|---|
| 90 nm | Sputtering | 400 ˚C | TiN/$Al_2O_3$ (001) | 0.86 nm [24] |
| 20 nm | MBE | 1150 ˚C | AlN/ $Al_2O_3$ (001) | 0.80 nm [25] |
| 80 nm | Sputtering | 400 ˚C | Si (100) | 0.50 nm [29] |
| 20 nm | MBE | 800 ˚C | 6H-SiC (001) | 0.46 nm [26] |
| 28 nm | ALD | 300 ˚C | $HfO_2$/$SiO_2$/Si (110) | 0.31 nm [27] |
| 30 nm | PLD | 600 ˚C | MgO | 0.17 nm [28] |
| 5 nm | MBE | 800 ˚C | 6H-SiC (001) | 0.15 nm [12] |
| 30 nm | Sputtering | 25 ˚C | TiN/ $Al_2O_3$ (001) | 0.12 nm [This work] |

[a] Table footnote.

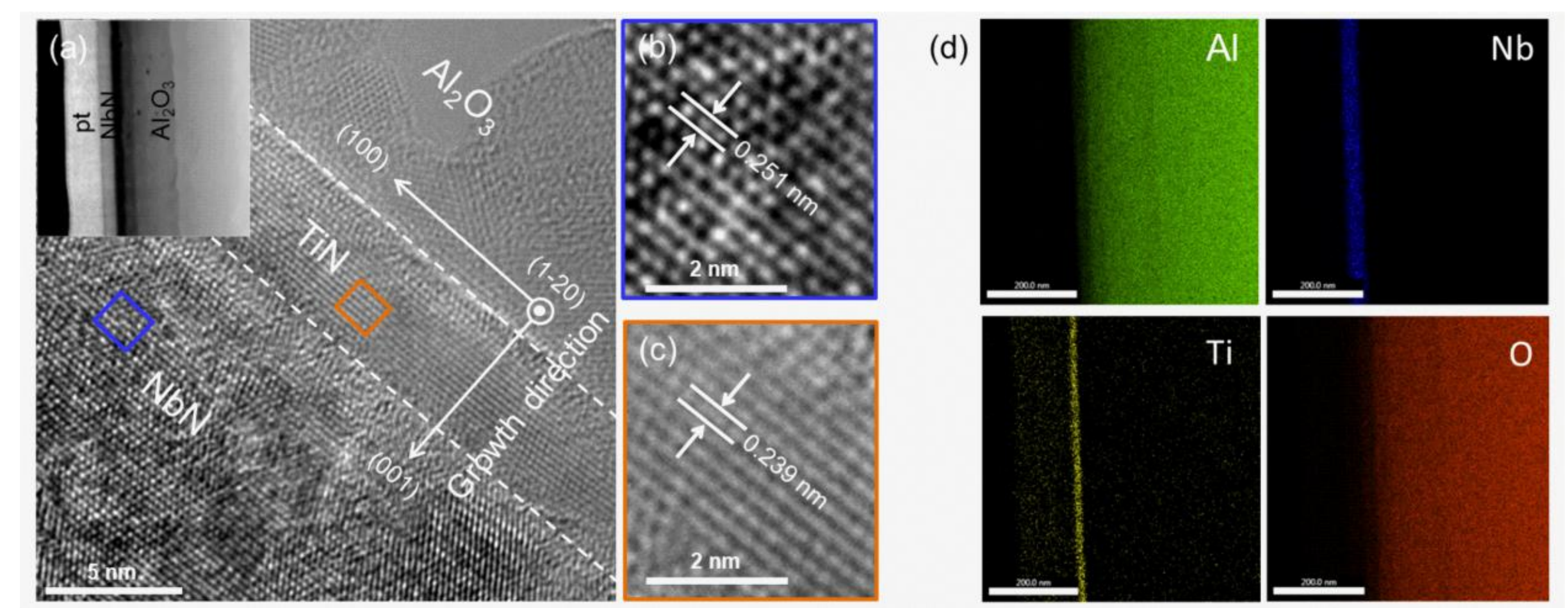

**Figure 5.** (a) Cross-sectional TEM image of sample E, the inset shows a lower magnification view, while the higher magnification image reveals the atomic arrangements of NbN and TiN and their sharp interface. (b, c) Close-up view of the marked regions, showing the single crystallinity of NbN and TiN with interplanar spacings 0.251 nm and 0.239 nm, respectively. (d) Elemental mapping of Al, Nb, Ti and O, indicating no oxygen diffusion into the NbN layer.

A further reduction in RMS roughness (0.212 nm for sample D) is observed as the TiN thickness is reduced. However, sample E shows the lowest roughness among all samples, 0.122 nm, even though the TiN layer in this case is relaxed. To the best of our knowledge, this is the lowest reported roughness for NbN, achieved without the use of any sophisticated deposition tool such as MBE, or by including high-temperature processing steps. Table 2 compares the roughness values of the obtained films with those reported earlier.[12, 24-29]

To further investigate the crystalline quality and interfaces of the deposited films, HRTEM images were recorded from sample E, and are shown in Fig. 5(a). A sharp interface between the NbN and TiN films, as well as between TiN and sapphire is observed. No interfacial layer is seen in either the NbN-TiN or the TiN-sapphire interfaces, which is due to close lattice match between the respective pair of layers. However, when NbN is grown directly on Si, an amorphous interlayer is present at the interface.[30] Figure 5(a) also shows that both NbN and TiN are crystalline, and are epitaxially stacked along (001) direction of sapphire. The lower magnification image, shown in the inset of Fig. 1(a), allows the thickness of the NbN layer to be measured as 36.269 nm. The higher magnification images, captured at the regions shown by rectangular boxes in Fig. 5(a), are presented in Fig. 5(b) and Fig. 5(c). The interplanar spacing, measured from these images are 0.251 nm and 0.239 nm, which correspond to the (111) planes of NbN and TiN, respectively. Elemental mapping of Al, Nb, Ti, and O, performed using TEM energy dispersive X-ray (EDX) analysis, reveals composition homogeneity of the layers and the absence of interdiffusion of any element during growth. Furthermore, elemental mapping of oxygen shows a uniform distribution in the sapphire region, while it is not detected in either the NbN or the TiN region, or at their interface.

As micro strain, Nb-N stoichiometry, and oxygen incorporation in the NbN film lead to a reduction in the superconducting critical temperature ($T_c$), XPS measurements for the Nb 3d, Nb 3p and N 1s core levels were performed to quantitatively evaluate the strain induced modification of bond strength, stoichiometry and the chemical state of oxygen in the film. Fig. 6(a) and Fig. 6(b) show the spectra for Nb 3d for sample A and sample D, respectively. The chemical compounds were determined by fitting the spectra with gaussian curves. The peak located at binding energies of 202.9, 204.5, and 206.7 eV are attributed to metallic Nb, Nb-$N_{(3d5/2)}$ and Nb-$N_{(3d3/2)}$, respectively. On the other hand, the peaks at 203.7, 207.9, and 209.5 eV correspond to Nb-O, $Nb_2O_{5(3d5/2)}$ and $Nb_2O_{5(3d3/2)}$, respectively, while the peaks observed at 205.5, 208.9 eV are assigned to Nb-N-$O_{(3d5/2)}$ and Nb-N-$O_{(3d3/2)}$, respectively. To check whether nitrogen content remains same in each sample, the areas under the N1s and Nb 3p spectra were estimated from Fig. 6(c), and their ratio was compared. For all samples, the ratio varies between 0.11-0.13, indicating approximately the same nitrogen content. Since the layers were deposited at the same temperature (room-temperature), the Nb-N stoichiometry is expected to remain similar for all samples. It is to be noticed that the binding energy of N 1s shifts with TiN thickness, which might be due to the variation in bond strengths caused by strain. To investigate the influence of strain on the strength of the Nb-N bond, the binding energy of $Nb_{(3d5/2)}$ for Nb-N and its difference from the N1s binding energy were plotted, as shown in Fig. 6(d). Although the binding energy of Nb $3d_{5/2}$ decreases with TiN thickness reduction, the difference $\Delta = N\,1s - Nb\,3d5/2$ increases. This suggests that the strength of Nb-N bond increases with the reduction of TiN thickness. It also indicates that there is micro strain in the NbN film, which increases as the TiN thickness reduces. However, the magnitude of the strain is too small to be detected by XRD measurements.

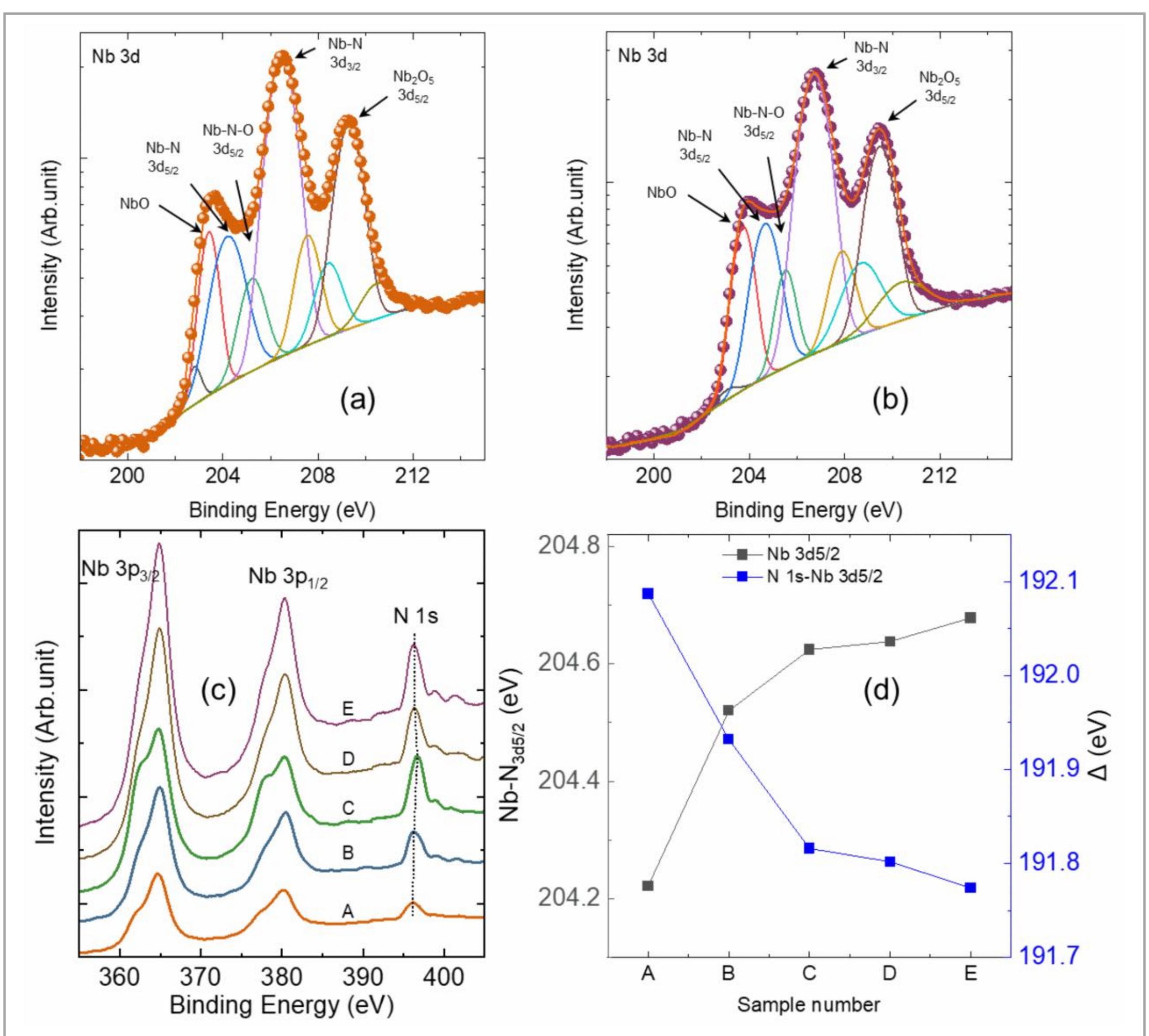

**Figure 6.** (a, b) Nb 3d spectra of sample A and sample D, showing the different elemental components contributing to the spectra, obtained from fitting the curves with Gaussian functions. (c) Nb 3p and N1s core-level spectra; the dotted line represents the shift of the N 1s peak. (d) Binding energy of Nb-N 3d5/2 and its difference from N 1s for all samples.

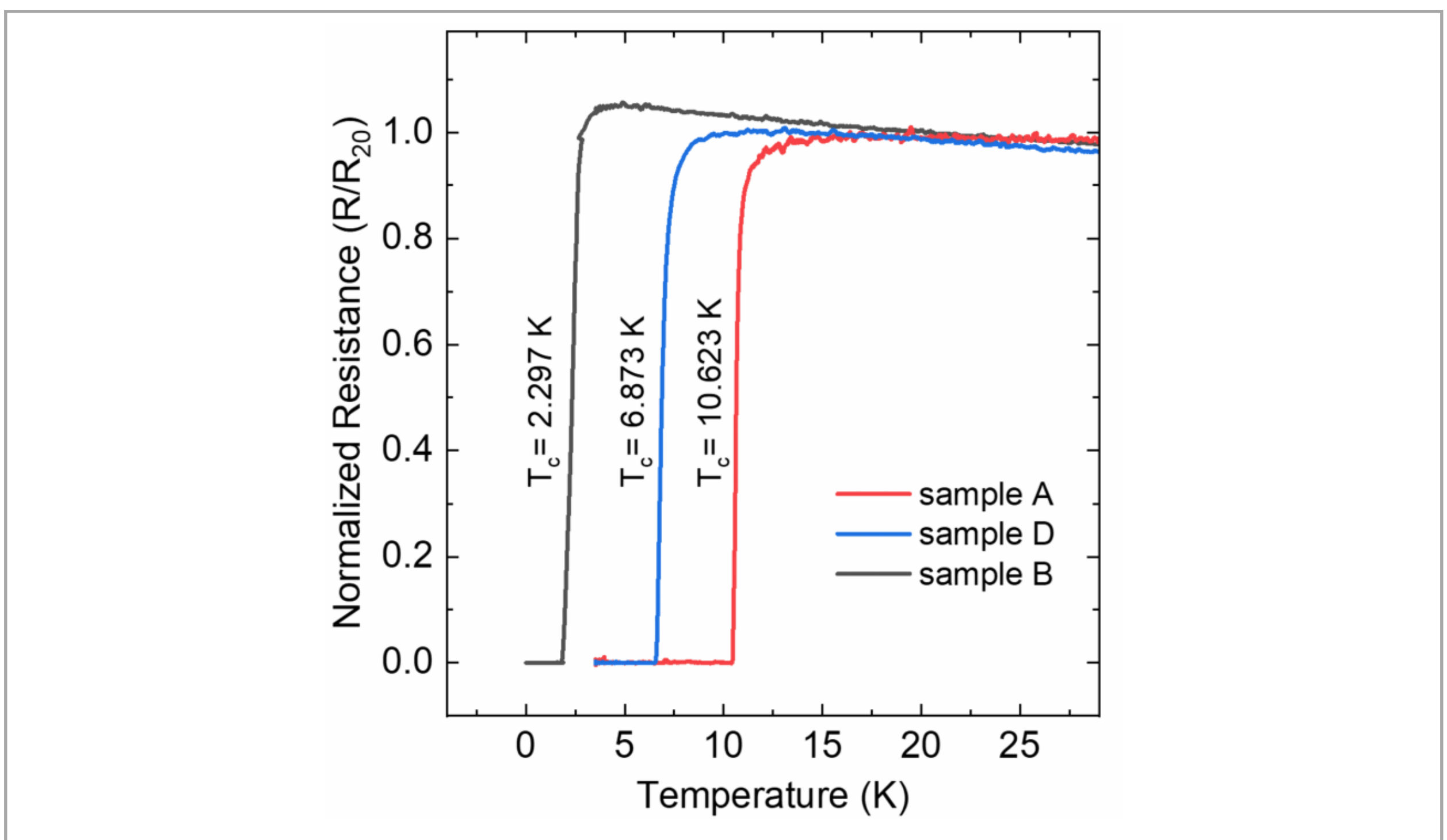

**Figure 7.** Temperature-dependent four-probe measurement of sample A, sample B, and sample D, showing a shift of $T_c$ to lower temperatures for sample B and sample D compared to sample A.

The previous experiments reveal that the NbN deposited on TiN buffer is highly crystalline and phase-pure and exhibits a sharp interface. To estimate the superconducting temperature, the four-probe resistance measurements were performed on those samples, in the van der-Pauw configuration. The variation of resistance with temperature is plotted and shown in Fig. 7. The measured $T_c$ values for sample A, sample D, and sample B are 10.623 K, 6.873 K, and 2.297 K, respectively. $T_c$ was determined as the temperature at which the resistance drops to 50% of its normal-state value. It can be clearly observed that the NbN film grown without a TiN buffer layer shows a higher $T_c$ than the NbN grown with a TiN buffer layer. Despite the high quality of the NbN film and the NbN/TiN interface, a reduction in $T_c$ is observed, when a TiN buffer layer is used. This cannot be solely attributed to strain, as the maximum change in $T_c$ due to strain does not exceed 1 K. However, we observe a deviation of 3.750 K between sample A and sample D, and 8.326 K between sample A and sample B. The primary reason for this is the proximity effect, which leads to the leakage of Cooper-pairs from the NbN layer into the TiN buffer layer. The leakage becomes more significant when a thicker TiN layer is used. In this case, NbN and TiN form a bilayer system, in which NbN has a higher $T_c$ compared to TiN. For the quantitative analysis of the leakage of cooper-pairs, we have used McMillan's proximity effect model. According to this model, the $T_c$ of NbN-TiN bilayer system can be described by the following equation[31, 32]:

$$ln\frac{{T_c}^0}{T_c} = \frac{\tau_{TiN}}{\tau_{TiN}+\tau_{NbN}}\left[\phi\left(\frac{1}{2}+\frac{(\tau_{TiN}+\tau_{NbN})\eta}{k_B T_c \tau_{TiN}\tau_{NbN}}\right) - \phi\left(\frac{1}{2}\right) - \ln\sqrt{1+(\frac{\tau_{TiN}+\tau_{NbN}}{\tau_{TiN}\tau_{NbN}\omega_D})^2}\right] \quad (1)$$

Here, ${T_c}^0$ is the superconducting transition temperature of the NbN layer alone, while $\phi(x)$ and $\omega_D$ denote the digamma function and Debye frequency, respectively. The terms $\eta$ and $k_B$ are the Planck constant and Boltzmann constant, respectively. The other two terms, $\tau_{TiN}$ and $\tau_{NbN}$, are related to the properties of the respective layers and their interface. These two terms are expressed as follows

$$\tau_{NbN} = \pi\frac{d_{NbN}}{V_{NbN}}\rho_{int}$$

$$\tau_{TiN} = 2\pi\frac{V_{TiN}d_{TiN}}{{V_{NbN}}^2}\rho_{int} \quad (2)$$

Where $d_{NbN,TiN}$ and $V_{NbN,TiN}$ are the thickness and Fermi velocity of the respective layers, and $\rho_{int}$ is the interface transparency parameter. Considering the Fermi velocities of $0.7 \times 10^6\ m/s$ for NbN [33] and $1.5 \times 10^5\ m/s$ for TiN [34], and using equations (1) and (2), the interface transparency parameter was estimated to be 0.6 for sample D, and 0.45 for sample B. These values indicate that Cooper-pairs leak through the interface into the TiN layer quite efficiently. Using equation (2), the ratio of the effective coupling parameter $\frac{\tau_{TiN}}{\tau_{NbN}}$ was evaluated, which is 0.14 for sample D, and 0.87 for sample B. This implies that in the former case, $\tau_{NbN} \gg \tau_{TiN}$, superconductivity is dominant in NbN. However, in the latter case, $\tau_{NbN} \approx \tau_{TiN}$, superconductivity spreads across both layers, resulting in a decreased $T_c$. Since in this case, the superconductivity is dominated by the TiN layer, a significant reduction in $T_c$ is observed when the TiN thickness is increased.

## Conclusions:

In conclusion, we have deposited single crystalline δ-NbN on TiN/$Al_2O_3$(001) by room-temperature sputtering. The films are not only phase-pure but also exhibit a ultra-low roughness, which is attributed to controlled strain relaxation facilitated by the TiN buffer layer. It is observed that the thickness of the TiN layer plays a crucial role in realizing the high crystallinity NbN. Further analysis reveals a sharp interface and absence of any oxide interlayer between the NbN and TiN layers or between the TiN layer and sapphire substrate. The temperature dependent four probe measurements indicate that $T_c$ decreases when a TiN buffer layer is used, which is due to leakage of Cooper pairs from the NbN to the TiN layer. Moreover, our in-depth investigations show that the leakage of Cooper pairs increases with the increase of the TiN layer thickness, which in turn leads to a further reduction in $T_c$. Therefore, this work establishes the feasibility of growing high-quality single crystalline δ-NbN on TiN-buffered sapphire substrate, by low-cost room temperature sputtering, which is beneficial for fabrication of Josephson junction-based devices, kinetic inductors and single photon detectors.

## Acknowledgements:

All authors gratefully acknowledge the financial support from the Quantum Information Technologies with Superconducting Devices and Quantum Dots (Project no- RD/0120-DSTIC01-001), and the Nano-electronics Network for Research and Applications (NNetRA) (Project Code No. RD/0520-IRNTRS0-001) of the Govt. of India (GoI). The IIT Bombay Nanofabrication Facility (IITBNF) is acknowledged for technical support in execution of the work.